\documentclass[aps,twocolumn,superscriptaddress,showpacs,letter,nofootinbib]{revtex4-1} 
\usepackage[utf8]{inputenc}
\usepackage{graphicx}
\usepackage{dcolumn}
\usepackage{bm}
\usepackage{cancel}
\usepackage{color}             
\usepackage{epsfig}
\usepackage{amssymb}
\usepackage{amsmath}
\usepackage{multirow}
\usepackage{hyperref}
\usepackage{cleveref}
\usepackage{lineno,color}
\usepackage{epstopdf}

\begin{document}

\title{Unitarization Schemes for High-Energy Elastic Scattering}

\author{E.~G.~S.~Luna}
\email{luna@if.ufrgs.br}
\affiliation{Instituto de F\'isica, Universidade Federal do Rio Grande do Sul, Caixa Postal 15051, 91501-970, Porto Alegre, Rio Grande do Sul, Brazil}
 

\begin{abstract}

We examine key features of the eikonal and $U$-matrix unitarization schemes and apply both frameworks to investigate the high-energy behavior of the elastic scattering amplitude, considering input amplitudes whose asymptotic behavior is driven exclusively by Pomeron exchange. Particular attention is devoted to assessing how the extracted Pomeron parameters depend on the assumed trajectory form. In this context, we compare linear and nonlinear Pomeron trajectories, evaluating their impact on the phenomenological description of the data and on the stability of the fitted parameters. 

\end{abstract}


\maketitle

\section{Introduction}

In a previous study \cite{mlp001}, we investigated the high-energy behavior of the elastic scattering amplitude within two unitarization schemes: the eikonal representation and the $U$-matrix approach. Part of that analysis was devoted specifically to the Pomeron sector. The fits were performed using two distinct data ensembles, one based on measurements from the ATLAS Collaboration and the other from TOTEM, comprising total cross sections, the $\rho$ parameter, and $pp$ differential cross-section data at LHC energies, specifically $\sqrt{s} = 7$, 8, and 13 TeV, in the interval $|t| \leq 0.1$ GeV$^{2}$.

In the present work, we adopt the same theoretical framework and statistical methodology employed in \cite{mlp001}, but with a different focus. Rather than discussing the tension between the differential cross-section results reported by the TOTEM and ATLAS Collaborations \cite{lkmr01,bopsin01,petrov01,petrov02}, we aim to examine the sensitivity of the extracted Pomeron parameters to the choice of its Regge trajectory, concentrating for the moment exclusively on the ATLAS dataset. In particular, we compare the outcomes obtained with a linear trajectory to those derived from a nonlinear parametrization to assess the extent to which the fits depend on this modeling assumption.

The outline of this paper is as follows. In Sec. II, we present a detailed discussion of the unitarization schemes, emphasizing their formal structure and the role of the impact-parameter representation. In Sec. III, we introduce the functional forms adopted for the Born (input) amplitudes, specifying the parametrizations for the Pomeron contribution and the corresponding Regge trajectories. Finally, in Sec. IV, we present our results and draw our conclusions.

\section{Unitarization Schemes}

In quantum scattering theory, the dynamics is encoded in a unitary $S$-matrix operator,
\begin{eqnarray}
S^{\dagger} S = S S^{\dagger} = \mathbf{1}.
\end{eqnarray}

In a partial-wave decomposition, elastic unitarity implies
\begin{eqnarray}
S_{\ell} = e^{2 i \delta_{\ell}},
\end{eqnarray}
where $\delta_{\ell}$ is the phase shift. An equivalent parametrization is obtained in terms of the reaction matrix (or $K$-matrix) \cite{landau01,taylor01},
\begin{eqnarray}
S_{\ell} = \frac{1 + i K_{\ell}}{1 - i K_{\ell}},
\end{eqnarray}
with $K_{\ell}$ Hermitian below inelastic thresholds. The two representations are related by
\begin{eqnarray}
K_{\ell} = \tan \delta_{\ell}.
\end{eqnarray}

In this strict sense, the two forms are simply different parametrizations of the same elastic amplitude.
More generally, any unitary operator $S$ (which does not have $-1$ as an eigenvalue in its spectrum) can be expressed in terms of a Hermitian operator $T$ through a Cayley transform,
\begin{eqnarray}
S = \frac{1 + i T}{1 - i T}.
\end{eqnarray}
Thus, once the full physical $S$ is known, the exponential and Cayley forms are related by a change of variables. This mathematical equivalence, however, does not imply dynamical equivalence when different functional embeddings are chosen for phenomenological purposes.

At high energies, it is convenient to formulate the scattering process in impact-parameter space. In this representation, the $S$-matrix density $S(s,b)$ obeys the unitarity bound
\begin{eqnarray}
|S(s,b)| \le 1,
\label{bound01}  
\end{eqnarray}
with equality holding in the limit of purely elastic scattering. Introducing the elastic profile function $H(s,b)$ through
\begin{eqnarray}
S(s,b) = 1 + i\, H(s,b),
\end{eqnarray}
the unitarity condition can be written in local form as
\begin{eqnarray}
2\,\textnormal{Im}\, H(s,b) = |H(s,b)|^{2} + G_{in}(s,b),
\end{eqnarray}
where $G_{\rm inel}(s,b) \ge 0$ denotes the inelastic overlap function. This relation makes explicit the interplay between elastic and inelastic contributions at fixed impact parameter $b$. The unitarized scattering amplitude in momentum space, ${\cal A}(s,t)$, is then obtained from the profile function $H(s,b)$ through the inverse Fourier–Bessel transform
\begin{eqnarray}
{\cal A}(s,t) = s \int_{0}^{\infty}b\, db\, J_{0}( b\sqrt{-t}) \, H(s,b) .
\label{scatter007}
\end{eqnarray}

Unitarization, however, is not uniquely determined by the unitarity condition alone. This condition defines a domain (the unit disk in the complex plane), but it does not prescribe how a given Born input should be mapped into that domain. Rather, it amounts to selecting a specific holomorphic mapping that sends a complex input function, typically defined in the upper half-plane, into the unit disk. In impact-parameter space, both the exponential (eikonal-type) representation,
\begin{eqnarray}
S(s,b) = e^{i\chi(s,b)},
\end{eqnarray}
and the Cayley (or $U$-matrix-type) representation,
\begin{eqnarray}
S(s,b) = \frac{1 + i U(s,b)}{1 - i U(s,b)},
\end{eqnarray}
are examples of such mappings. They both ensure unitarity by construction, yet they correspond to different analytic embeddings of the input amplitude. The formal equivalence between exponential and Cayley representations holds only if one starts from the exact physical $S(s,b)$ and performs a change of variables, for instance, through the relation $U(s,b)=\tan (\chi(s,b)/2)$. In that case, the two parametrizations describe the same physical object and differ merely by a redefinition of variables. This is completely analogous to the equivalence between phase-shift and reaction-matrix descriptions in quantum scattering theory.

In practical phenomenological applications to high-energy hadronic scattering, however, the situation is fundamentally different. One does not begin with the exact $S(s,b)$. Instead, one constructs a model Born amplitude, typically associated with Regge exchanges or Pomeron contributions, and embeds it in a chosen functional form that enforces unitarity. The exponential and Cayley schemes, therefore, do not act on the same function; they act on the same Born input but generate different unitary amplitudes.

This distinction has important dynamical consequences. Although both schemes resum multiple scatterings, they do so with different combinatorial weights. The eikonal representation generates an exponential (Poisson-like) resummation pattern, whereas the $U$-matrix scheme yields a rational (geometric-type) resummation. As a result, the hierarchy of higher-order multiple interactions differs, affecting the approach to saturation and the relative strength of elastic versus inelastic channels.

In particular, the two schemes exhibit different asymptotic behaviors in the strong-scattering regime. The standard eikonal scheme leads to the so-called black-disk limit, characterized by $S(s,b) \to 0$. In contrast, the $U$-matrix scheme allows for the possibility of reflective scattering, where $S(s,b)$ may become negative and approach $-1$, corresponding to a qualitatively different saturation mechanism.

Therefore, while both schemes satisfy the same unitarity constraint and can be formally related at the level of the exact $S$-matrix, they represent genuinely distinct dynamical implementations when applied to modeled Born inputs. Their differences are not merely algebraic but structural, and they may lead to observable consequences, particularly in quantities sensitive to the balance between elastic and inelastic contributions.

The construction of unitarized scattering amplitudes begins with the specification of a Born (or input) term, namely a model amplitude in momentum space, ${\cal F}(s,t)$. Its crossing-even and crossing-odd components are defined in the usual way as
\begin{eqnarray}
{\cal F}^{\pm}(s,t) = \frac{1}{2} \left[ {\cal F}^{pp}(s,t)\pm {\cal F}^{\bar{p}p}(s,t)\right].
\end{eqnarray}
The corresponding Born amplitudes in impact-parameter space are
\begin{eqnarray}
\chi^{\pm}(s,b) = \frac{1}{s}\int_{0}^{\infty} q\, dq\, J_{0}(bq)\, {\cal F}^{\pm}(s,-q^{2}),
\end{eqnarray}
where $-q^{2}=t$. Unitarization schemes are then obtained through the construction of an amplitude $H(s,b)$ in terms of the functions $\chi(s,b)$ \cite{mlp001,lr01,Oueslati01,Oueslati02,Oueslati03}. In the eikonal scheme (Es), one has
\begin{widetext}
\begin{eqnarray}
H_{[Es]}(s,b) = i \left[ 1 - e^{i\chi (s,b)}  \right] = -i\sum_{n=1}^\infty \frac{[i\chi(s,b)]^{n}}{n!} = i\sum_{n=1}^\infty C_{n}^{eik} \cdot (-1)^{n-1} [\Omega(s,b)]^{n} = i\left[ \frac{\Omega}{2} - \frac{\Omega^{2}}{8} + \frac{\Omega^{3}}{48} + ... \right] ,
\label{eikrelation01}  
\end{eqnarray}
\end{widetext}
where $C_{n}^{eik} = 2^{-n}/n!$, $J_0(x)$ is the Bessel function of the first kind, and 
$\Omega(s,b)\equiv -2i\chi(s,b)$ is the opacity of $pp$ interaction.

The $U$-matrix unitarization (Us) leads to the relation
\begin{widetext}
\begin{eqnarray}
H_{[Us]}(s,b) = \frac{\hat{\chi}(s,b)}{1-i\hat{\chi}(s,b)/2} = -2i \sum_{n=1}^\infty \frac{[i\hat{\chi}(s,b)]^{n}}{2^{n}} =
  i\sum_{n=1}^\infty C_{n}^{U} \cdot (-1)^{n-1} [\hat{\Omega}(s,b)]^{n} = i\left[ \frac{\Omega}{2} - \frac{\Omega^{2}}{8} + \frac{\Omega^{3}}{32} + ... \right] ,
\label{eikrelation02}  
\end{eqnarray}
\end{widetext}
where $C_{n}^{U} = 2/4^{n}$, and $\hat{\Omega}(s,b)\equiv -2i\hat{\chi}(s,b)$ is the respective opacity.
In the expansion over the $\Omega$ or $\hat{\Omega}$ powers each $\Omega^n$ or $\hat{\Omega}^n$ term corresponds to the exchange of $n$ Pomerons. We can observe that the first two terms in expansions (\ref{eikrelation01}) and (\ref{eikrelation02}) (with $\Omega = {\Bbb P} = \hat{\Omega} $) are identical.

\section{The Pomeron Born amplitude}

The input amplitudes ${\cal F}_{i}(s,t)$ are interpreted as single Reggeon-exchange contributions at the Born level. For each exchanged trajectory, the corresponding amplitude is written as
\begin{eqnarray}
{\cal F}_{i}(s,t) = \beta_{i}^{2}(t)\eta_{i}(t)\left( \frac{s}{s_{0}} \right)^{\alpha_{i}(t)} ,
\label{equation05_new}
\end{eqnarray}
with $i=-, +, \Bbb P$. In this expression, $\beta_{i}(t)$ denotes the proton–Reggeon vertex function, $\eta_{i}(t)$ is the signature factor, $\alpha_{i}(t)$ represents the Regge trajectory, and the fixed energy scale is $s_{0} \equiv 1$ GeV$^{2}$.

The amplitudes ${\cal F}_{-}(s,t)$ and ${\cal F}_{+}(s,t)$ describe the exchange of secondary Reggeons with negative and positive charge conjugation, respectively. Specifically, ${\cal F}_{-}$ accounts for $C=-1$ exchanges (such as $\omega$ and $\rho$), whereas ${\cal F}_{+}$ corresponds to $C=+1$ exchanges (notably $a_{2}$ and $f_{2}$). The term ${\cal F}_{\Bbb P}(s,t)$ corresponds to the $C=+1$ Pomeron exchange, which dominates the high-energy behavior of the scattering amplitude.

For odd-signature trajectories, the signature factor takes the form
\begin{eqnarray}
\eta_{i}(t) = -i, e^{-i\frac{\pi}{2}\alpha_{i}(t)},
\end{eqnarray}
whereas even-signature trajectories are described by
\begin{eqnarray}
\eta_{i}(t) = -e^{-i\frac{\pi}{2}\alpha_{i}(t)}.
\end{eqnarray}

Reggeons with positive charge conjugation are modeled with an exponential proton–Reggeon vertex,
\begin{eqnarray}
\beta_{+}(t)=\beta_{+}(0)\exp \left( \frac{r_{+} t}{2} \right),
\end{eqnarray}
and are assumed to follow linear trajectories of the form
\begin{eqnarray}
\alpha_{+}(t) = 1 - \eta_{+} + \alpha^{\prime}{+} t.
\end{eqnarray}
Analogous parametrizations are adopted for the $C=-1$ secondary Reggeons, characterized by the parameters $\beta_{-}(0)$, $r_{-}$, $\eta_{-}$, and $\alpha^{\prime}_{-}$.

We have considered two distinct parametrizations for the Pomeron trajectory. The first one, which defines our ``{\bf Model I}'', is specified by
\begin{eqnarray}
\alpha_{\Bbb P}(t) = 1+\epsilon + \alpha^{\prime}_{\Bbb P} t .
\end{eqnarray}

The second Pomeron trajectory, referred to as ``{\bf Model II}'', has the nonlinear form \cite{lkmr01,kmr001,lrk001,broilo001}
\begin{eqnarray}
\alpha_{\Bbb P}(t) = \alpha_{\Bbb P}(0) + \alpha^{\prime}_{\Bbb P} t + \frac{m_{\pi}^{2}}{32\pi^{3}}\,
h(\tau) ,
\label{pomnlin}
\end{eqnarray}
where $\alpha_{\Bbb P}(0) = 1 + \epsilon$ and
\begin{eqnarray}
h (\tau) &=& -\frac{4}{\tau}\, F_{\pi}^{2}(t) \left[  2\tau - (1+\tau)^{3/2} \ln \left( \frac{\sqrt{1+\tau}+1}{\sqrt{1+\tau}-1} \right)
\right. \nonumber \\
& & + \left. \ln \left( \frac{m^{2}}{m_{\pi}^{2}} \right) \right] .
\label{nonlinear01}
\end{eqnarray}
Here, $\epsilon$ is assumed to be positive, $\tau$ is defined as $4m_{\pi}^{2}/|t|$, with $m=1$ GeV setting the hadronic scale and $m_{\pi}=139.6$ MeV denoting the pion mass. In Eq. (\ref{nonlinear01}), the function $F_{\pi}(t)$ represents the pion–Pomeron vertex form factor. We adopt a standard single-pole parametrization, $F_{\pi}(t)=\beta_{\pi}/(1-t/a_{1})$, where $\beta_{\pi}$ determines the strength of the pion–Pomeron coupling. Its normalization is fixed according to the additive quark model prescription, $\beta_{\pi}/\beta_{I\!\!P}(0)=2/3$.
The additional contribution appearing as the third term on the right-hand side of Eq. (\ref{pomnlin}) accounts for pion-loop corrections, arising from $t$-channel unitarity.

The proton–Pomeron vertex is parametrized as
\begin{eqnarray}
\beta_{\Bbb P}(t)=\beta_{\Bbb P}(0)\exp \left( \frac{r_{\Bbb P}t}{2} \right) .
\label{vertex01}
\end{eqnarray}

In both unitarization schemes, the full opacity in impact-parameter space is constructed as the sum of all exchange contributions,
\begin{eqnarray}
\Omega^{pp}_{\bar{p}p}(s,b) = \Omega_{\Bbb P}(s,b)+\Omega_{+}(s,b)\pm\Omega_{-}(s,b).
\end{eqnarray}

The observables are expressed in terms of the scattering amplitude ${\cal A}^{pp}_{\bar{p}p}(s,t)$: the total cross section is given by
\begin{eqnarray}
\sigma^{pp, \bar{p}p}_{tot}(s)=\frac{4\pi}{s}\, \textnormal{Im}\, {\cal A}^{pp}_{\bar{p}p}(s,t=0) ,
\end{eqnarray}
the elastic differential cross section by
\begin{eqnarray}
\frac{d\sigma^{pp, \bar{p}p}}{dt}(s,t)=\frac{\pi}{s^{2}}\, \left| {\cal A}^{pp}_{\bar{p}p}(s,t) \right|^{2} ,
\end{eqnarray}
and the $\rho$ parameter by
\begin{eqnarray}
\rho^{pp, \bar{p}p}(s)=\frac{\textnormal{Re}\, {\cal A}^{pp}_{\bar{p}p}(s,t=0)}{\textnormal{Im}\, {\cal A}^{pp}_{\bar{p}p}(s,t=0)} .
\end{eqnarray}

\begin{table*}
\centering
\caption{The Pomeron parameter values extracted from global fits to the selected data ensemble within the eikonal and $U$-matrix unitarization schemes.}
\begin{ruledtabular}
\begin{tabular}{cccccc}
 & \multicolumn{2}{c}{Eikonal unitarization} & \multicolumn{2}{c}{$U$-matrix unitarization} \\
\cline{2-3} \cline{4-5}  \\ [-0.3cm]
 & {Model I}  & {Model II} & {Model I} & {Model II}  \\   
 \hline \\ [-0.3cm]
$\beta_{\Bbb P}(0)$ & 2.15$\pm$0.13 & 2.154$\pm$0.063 & 2.25$\pm$0.11 & 2.271$\pm$0.075 \\
$\epsilon$ & 0.1021$\pm$0.0063 & 0.1014$\pm$0.0033 & 0.0930$\pm$0.0054 & 0.0911$\pm$0.0037 \\
$\alpha^{\prime}_{I\!\!P}$ (GeV$^{-2}$) & 0.297$\pm$0.094 & 0.2938$\pm$0.0022 & 0.418$\pm$0.065 & 0.4425$\pm$0.0085 \\
$r_{\Bbb P}$ (GeV$^{-2}$) & 2.4$\pm$1.7 & 2.375$\pm$0.019 & 0.60$\pm$0.52 & 0.1051$\pm$0.0061 \\
\hline \\ [-0.3cm]
$\nu$ & 226  & 226 & 226 & 226 \\
$\chi^{2}/\nu$ & 0.86  & 0.86 & 0.85 & 0.85 
\end{tabular}
\end{ruledtabular}
\label{tab001}
\end{table*}

\begin{figure*}\label{fig007}
\begin{center}
\includegraphics[height=.5\textheight]{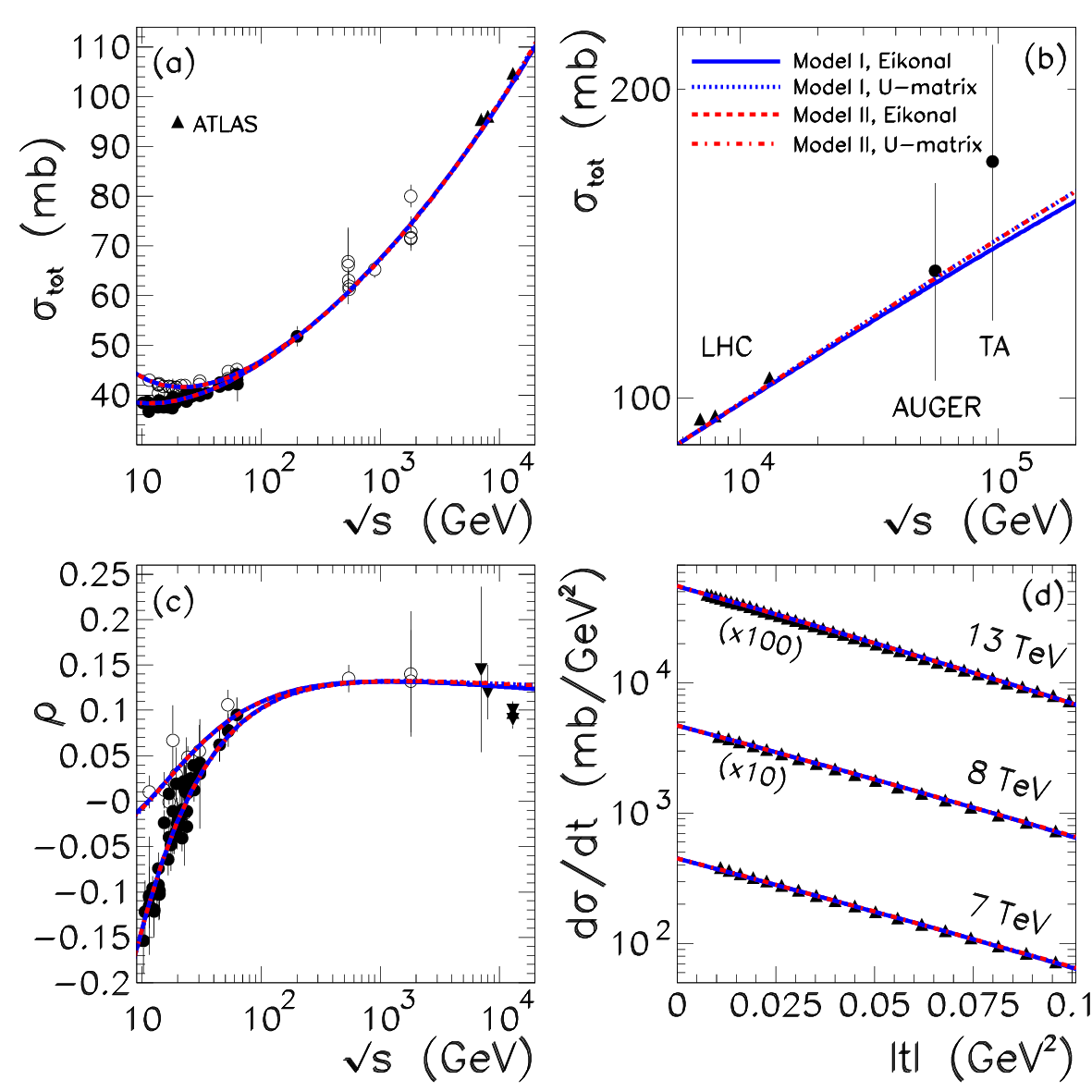}
\caption{(a),(b) Total cross section, (c) $\rho$ parameter, and (d) differential cross section for $pp$ and $\bar{p}p$ channels. Results obtained using eikonal amd $U$-matrix unitarizations.}
\end{center}
\end{figure*}

\section{Results and Conclusions}

We carry out simultaneous fits to the total cross section, $\sigma_{tot}^{pp,\bar{p}p}$, and to the $\rho^{pp,\bar{p}p}$ parameter, the ratio between the real and imaginary parts of the forward scattering amplitude, for energies above $\sqrt{s}=10$ GeV. In addition, we include $pp$ differential cross section data, $d\sigma^{pp}/dt$, measured at the LHC in the forward region, $|t|\leq 0.1$ GeV$^{2}$. The $\sigma_{tot}$ and $\rho$ datasets are taken from the compilation provided by the Particle Data Group \cite{pdg001}, while the differential cross-section data at 7, 8, and 13 TeV are from the ATLAS Collaboration \cite{atlas001,atlas002, ATLAS01}. For all datasets, statistical and systematic uncertainties are combined in quadrature.

The analysis is performed using the standard $\chi^{2}$ minimization procedure. The minimum value, $\chi^{2}_{min}$, is assumed to follow a $\chi^{2}$ distribution with $\nu$ degrees of freedom. Confidence regions are defined through the usual $\chi^{2}$ criterion; for eight free parameters, the 90\% confidence level corresponds to $\chi^{2}-\chi^{2}{min}=13.36$, assuming Gaussian uncertainties.

To reduce the number of adjustable parameters, the slopes of the secondary Reggeon linear trajectories are fixed at $\alpha^{\prime}_{+}=\alpha^{\prime}_{-}=0.9$ GeV$^{-2}$, consistent with the typical values extracted from Chew–Frautschi systematics. Likewise, the slope parameters entering the secondary-Reggeon form factors are set to $r_{+}=r_{-}=4.0$ GeV$^{-2}$. These quantities exhibit weak correlations with the Pomeron sector, and their adopted values agree with previous phenomenological analyses \cite{mlp001,lkmr01,bopsin01}, remaining stable across different unitarization schemes. Under these conditions, our discussion concentrates on the Pomeron parameters.

The Pomeron parameters obtained from the global fits to the selected data ensemble, within both the eikonal and $U$-matrix unitarization approaches, are listed in Table \ref{tab001}. The corresponding descriptions of the data are displayed in Figure 1, where $pp$ and $\bar{p}p$ results are compared for the two schemes. Panel (a) shows the total cross section, while panel (b) presents the same observable over an extended energy range. Panel (c) displays the $\rho$ parameter, and panel (d) the differential cross section in the forward region. For completeness, panel (b) also includes cosmic-ray estimates of the $pp$ total cross section, namely the AUGER measurement at $\sqrt{s}=57$ TeV \cite{auger} and the Telescope Array result at $\sqrt{s}=95$ TeV \cite{TA}.

The results obtained with the eikonal and $U$-matrix unitarization schemes agree with the previous analysis reported in Ref.\cite{mlp001}. The fitted parameter values exhibit only minor variations, and the corresponding $\chi^{2}/\nu$ values remain low in all cases, indicating that the overall description of the data is largely insensitive to the specific choice of Pomeron trajectory. Noticeable differences between the two unitarization schemes emerge only at very high energies, in the cosmic-ray domain, as illustrated in panel (b) of Figure 1.

It is also evident that, within the eikonal framework, Models I and II provide practically indistinguishable descriptions of the data: the curves representing the observables overlap to the extent that no visible differences can be discerned. The same behavior is observed in the $U$-matrix scheme, where both trajectory choices lead to nearly identical predictions across the fitted energy range.

The extracted values of $\beta_{\Bbb P}(0)$ are mutually compatible, within uncertainties, for all models and schemes considered. However, the $U$-matrix approach systematically yields slightly smaller values of the Pomeron intercept parameter $\epsilon$ compared to the eikonal case. The opposite trend is observed for the slope parameter $\alpha^{\prime}_{\Bbb P}$: the $U$-matrix fits tend to yield larger values than those obtained in the eikonal framework.

The most pronounced difference between the two schemes appears in the Pomeron slope parameter $r_{\Bbb P}$. In the eikonal approach, $r_{\Bbb P}$ remains stable, with values very similar between Models I and II. In contrast, within the $U$-matrix scheme, the fitted values of $r_{\Bbb P}$ are smaller, and in Model II, the value is nearly six times lower than in Model I.

From a statistical standpoint, however, all models and unitarization schemes describe the experimental data with comparable accuracy, as reflected by the nearly identical $\chi^{2}/\nu$ values. Nevertheless, none of the scenarios provides a satisfactory description of the $\rho^{pp}$ measurement at $\sqrt{s}=13$ TeV, which may signal the need for an additional $C=-1$ contribution, such as an Odderon term.

\section*{Acknowledgment}

The author is grateful to Roman Ryutin and the other members of the Organizing Committee for the kind invitation to participate in the XXXVIIth Workshop on High Energy Physics. We also thank Vladimir Petrov for raising stimulating and relevant physical questions, and for fostering an excellent environment for fruitful scientific discussions.

\end{document}